\input{psfig}
\documentstyle[aps,prc]{revtex}

\begin{document}


\title{Time-dependent properties of proton decay from crossing single-particle metastable states in deformed nuclei}

\author{
P. Talou${}^1$\thanks{E-mail: talou@cenbg.in2p3.fr}{\ },
N. Carjan${}^1$\thanks{E-mail: carjan@in2p3.fr}{\ } and
D. Strottman${}^2$\thanks{E-mail: dds@lanl.gov}
\\[1.5ex]
{\it ${}^1$CEN Bordeaux-Gradignan, F-33175 Gradignan Cedex, France} \\
{\it ${}^2$LANSCE and Theoretical Division, Los Alamos National Laboratory, Los Alamos, NM 87545, USA}
}

\maketitle

\begin{abstract}
A dynamical study of the decay of a metastable state by quantum tunneling through an anisotropic, non separable, two-dimensional potential barrier is performed by the numerical solution of the time-dependent Schr\"odinger equation. Initial quasi-stationary proton states are chosen in the framework of a deformed Woods-Saxon single-particle model. The decay of two sets of states corresponding to true and quasi level-crossing is studied and the evolution of their decay properties as a function of nuclear deformation is calculated around the crossing point. The results show that the investigation of the proton decay from metastable states in deformed nuclei can unambiguously distinguish between the two types of crossing and determine the structure of the nuclear states involved.
\end{abstract}





\section{Introduction}

The problem of the decay of a metastable state is of major importance in many different fields like nuclear chemistry, diffusion processes, spontaneous radioactivity, etc (for a review, see \cite{Jor86,Han90}). A metastable state, or quasi-stationary state, is defined as a state of local stability which decays with a rather large but finite lifetime toward a true stable minimum. When the temperature of the decaying system is low, quantum tunneling is the dominant decay process.

The one-dimensional tunneling process has been studied intensively from the very beginning of quantum mechanics and it is now quite thoroughly understood, at least within a stationary approach \cite{Gam28,Bri85}. On the other hand, little is known about multidimensional tunneling although some progress has been made through the last two decades \cite{Col77,Sch86,Wil86-87,Raz88}. Most of these approaches are restricted to special cases and cannot constitute a full and general framework for the study of multidimensional tunneling. \\
The recent, huge progress of computer performances makes possible the direct treatment of this complex quantum phenomenon through the numerical solution of the Time-Dependent Schr\"odinger Equation (TDSE) for well prepared initial quasi-stationary states, and hence the study of the dynamics of their decay. This procedure has been applied successfully to $\alpha$-decay \cite{Ser94} and fission \cite{Car94} in a one-dimensional model. A transient period at the very beginning of the decay process has been found that corresponds to deviations from the well-known exponential decay law \cite{Per80}. The relation between the asymptotic value of the time-dependent decay rate and the commonly used stationary value was also pointed out.

Recently, this method has been generalized to two dimensions to study the tunneling of quasi-stationary states through an anisotropic, non separable, potential barrier  \cite{Tal98}. Applied to proton emission from excited states in deformed nuclei, it has been shown that the escape directions chosen by the proton state depend mainly on its quantum numbers, i.e., on its spatial distribution, rather than on the features of the potential. The important role played by the distribution of the angular momentum and its variation in time for the determination of the proton decay rate have been also pointed out. Along with the experimental observation of nuclei beyond the proton dripline \cite{Hof96}, several stationary calculations for the half-lives of spherical \cite{Abe97} and deformed \cite{Bug89} ground state proton emitters have been reported.

It is well known that as a function of nuclear deformation, some single-particle levels cross each other \cite{Nil55}. Two distinct types of crossing can occur: 1) the quantum numbers (spin projection $\Lambda$ and parity $\pi$) of the crossing levels are the same but belong to two different oscillator shells ($\Delta N = 2$) ; in this case, the levels repel each other and their wave functions strongly mix near this quasi-crossing point. 2) the quantum numbers are different ; in this case, the two levels do not interact and we deal with a true crossing point \cite{BM75,She67}. \\
Therefore one roughly expects the proton decay properties of the two levels: 1) to be similar for quasi-crossing, the only difference coming from the small 'interaction' energy ; 2) to be very different for true crossing, although the two levels have exactly the same energy. This implies that the proton decay from excited states in deformed nuclei can represent a new tool for studying the level-crossing phenomenon and defining its type. Using the numerical, time-dependent approach mentioned above, the present paper offers a detailed study of the decay properties of single proton excited states for each of the two cases at and around the crossing deformations.

In the first part, the numerical formalism used for solving the TDSE in two dimensions is described as well as the physical quantities accessible with this approach. The deformed single-particle model used to obtain the proton quasi-stationary states inside the nucleus is described. In the second part, the time and deformation dependencies of the numerical results are presented and discussed. Finally, a summary, a conclusion and possible developments of this work are given.


\section{Formalism} \label{sec:Formalism}

\subsection{Time-dependent approach}

We are interested in describing the evolution of an individual proton interacting with an axially symmetric average potential created by the nucleons of the daughter nucleus. The total wavefunction representing the proton can be written in cylindrical coordinates as
\begin{eqnarray}
\psi (z,\rho,\phi, t) = f(z,\rho,t)e^{i\Lambda\phi}
\end{eqnarray}
where $\Lambda$ is the projection of the total angular momentum on the axis of symmetry ($z$-axis) and it is a good quantum number. Hence, the TDSE becomes 
\begin{eqnarray} \label{eq:TDSE}
i\hbar \frac{\partial}{\partial t}f(z,\rho,t) = \mathcal{H}(z,\rho)f(z,\rho,t)
\end{eqnarray}
where the hamiltonian $\mathcal{H}$ is
\begin{eqnarray} \label{eq:Hamilton}
\mathcal{H}(z,\rho) &=& -\frac{\hbar^2}{2\mu}\left[ \frac{1}{\rho}\frac{\partial}{\partial \rho}+ \frac{\partial ^2}{\partial \rho^2}+\frac{\partial^2}{\partial z^2}-\frac{\Lambda^2}{\rho ^2} \right] \\ \nonumber
&& + V_{pA}(z,\rho).
\end{eqnarray}
$V_{pA}$ is the potential representing the interaction between the individual proton and the remaining nucleons of the core daughter nucleus. 

In order to follow the time evolution of an initial state $\psi (\vec{r},t=0)$, the TDSE (\ref{eq:TDSE}) has been solved numerically within a finite spatial grid using a multistep predictor method called MSD2 \cite{Iit94}. This method presents a good compromise between stability, accuracy and efficiency for this type of numerical problems (linear, stationary potential). Within this approach, one has access to the wavefunction $\psi(\vec{r},t)$ at any time $t$ and then to the following physical quantities :

\begin{itemize}
\item[$\bullet$] {\bf Total tunneling probability} given by the fraction of the wavefunction located beyond the top of the two-dimensional potential barrier (potential ridge) at time $t$
\begin{eqnarray}
P_{tun}(t)=2\pi \int_{V_{out}}{|f(z,\rho,t)|^2 \rho d\rho dz};
\end{eqnarray}
\item[$\bullet$] {\bf Tunneling angular distribution} estimated in spherical coordinates $(r,\theta, \phi)$ and defined as
\begin{eqnarray}
P_{tun}(t,\theta)=\frac{dP_{tun}}{d\Omega}=\int_{r_{out}(\theta)}^{\infty}{|f(r,\theta,t)|^2r^2dr},
\end{eqnarray}
where $r_{out}(\theta)$ is the radial position of the potential ridge in the direction $\theta$. Note that
\begin{eqnarray}
2\pi \int_{0}^{\pi}{\frac{dP_{tun}}{d\Omega}sin\theta d\theta}=P_{tun};
\end{eqnarray}
\item[$\bullet$] {\bf Total decay rate} related to the total tunneling probability by
\begin{eqnarray}
\lambda(t)=\frac{1}{1-P_{tun}}\frac{dP_{tun}}{dt};
\end{eqnarray}
\item[$\bullet$] {\bf Mean value of angular momentum} defined as
\begin{eqnarray} \label{eq:l}
<L^2>(t) &=& <\psi(t)|L^2|\psi(t)> \nonumber \\
&\simeq & <l>(t)\left( <l>(t)+1 \right).
\end{eqnarray}
This quantity is important for the determination of the tunneling rate as well as of the angular momentum exchange during decay (non-spherical potential).
\end{itemize}

\subsection{Quasi-stationary states}

To accomplish the numerical scheme proposed above, knowledge of the initial metastable wavefunction is necessary. A quasi-stationary state is defined as an infinitesimally modified eigenstate of the hamiltonian. In a one-dimensional model, the two-potential formalism \cite{Gur87} can be used straightforwardly \cite{Ser94}. In our two dimensional approach, the numerical difference between the preparation of the initial quasi-stationary state as an expansion in an orthonormal basis (diagonalization) and the resolution of the TDSE on a discretized spatial grid has been proven to be sufficient to create quasi-stationary states without modifying the potential \cite{Tal98}.

In the case of an axially and reflexion symmetric potential, the quantum numbers labeling each single-particle proton state are $(\Lambda\pi)$ with $\Lambda$ the projection of the total angular momentum on the axis of symmetry and $\pi$ the parity. No spin-orbit term is present in our calculations.


\subsection{Single-particle potential}

The interaction between the emitted proton and the daughter nucleus is represented by the average potential $V_{pA}$. Obviously, this potential depends on the deformation of the parent nucleus whose shape has been described by Cassinian ovals \cite{Pas71}. Only one parameter $\varepsilon$ is sufficient to describe shapes from spheres ($\varepsilon=0$) to scission point configurations ($\varepsilon=1$).

Hence, $V_{pA}^\varepsilon$ can be written as
\begin{eqnarray}
V^\varepsilon_{pA} = V_\varepsilon^{nucl.} + V_\varepsilon^{Coul.}.
\end{eqnarray}
The nuclear part is a deformed Woods-Saxon potential
\begin{eqnarray}
V_\varepsilon^{nucl.}(z,\rho) = \frac{V_0}{1+exp(d_\varepsilon(z,\rho))}
\end{eqnarray}
where $d_\varepsilon(z,\rho)$ represents an approximation of the distance between the point $(z,\rho)$ and the deformed nuclear surface, $V_0$ the depth and $a$ the diffusivity of the nuclear potential. \\
The Coulomb part $V_\varepsilon^{Coul.}$ is obtained assuming that the nuclear volume is filled with a uniform charge distribution with a sharp edge.

\section{Numerical results}

As in our earlier work \cite{Tal98}, the numerical procedure described above has been applied to the hypothetical proton emission from excited states in a nucleus characterized by $Z=82$ and $A=208$. It is worth noting that, qualitatively, the results presented are independent of the choice of the decaying nucleus. While in this previous paper we concentrated on the general dependence of proton tunneling on time, nuclear deformation and quantum numbers, here we will analyze the behaviour of two selected sets of pair states around the true-crossing and quasi-crossing deformations, respectively. \\
The parameters used for the potential $V_{pA}$ have been chosen according to Ref. \cite{Ros68}. The depth $V_0$ has been modified in order to obtain states lying below the top of the two-dimensional barrier : $V_0 = -62.66$ MeV for the case of true-crossing, and $V_0 = -58.7$ MeV for the case of quasi-crossing.

\subsection{True-crossing states}

We studied the case of two single-particle states with different quantum numbers, $\psi_{2h} (\Lambda\pi=3-)$ and $\psi_{1k} (\Lambda\pi=3+)$\footnote{The indices ``2h'' and ``1k'' for the proton states have been chosen using the usual spectroscopic notation according to the angular momentum value of the corresponding spherical states ($l=5$ and $l=8$, respectively).}. \\ At the deformation $\varepsilon_{tc}=0.25$, the two corresponding energy levels cross, i.e., undergo an accidental degeneracy, $E_{2h}=E_{1k}$. This situation arises because of the multidimensionality of the problem.

\begin{figure}
\psfig{file=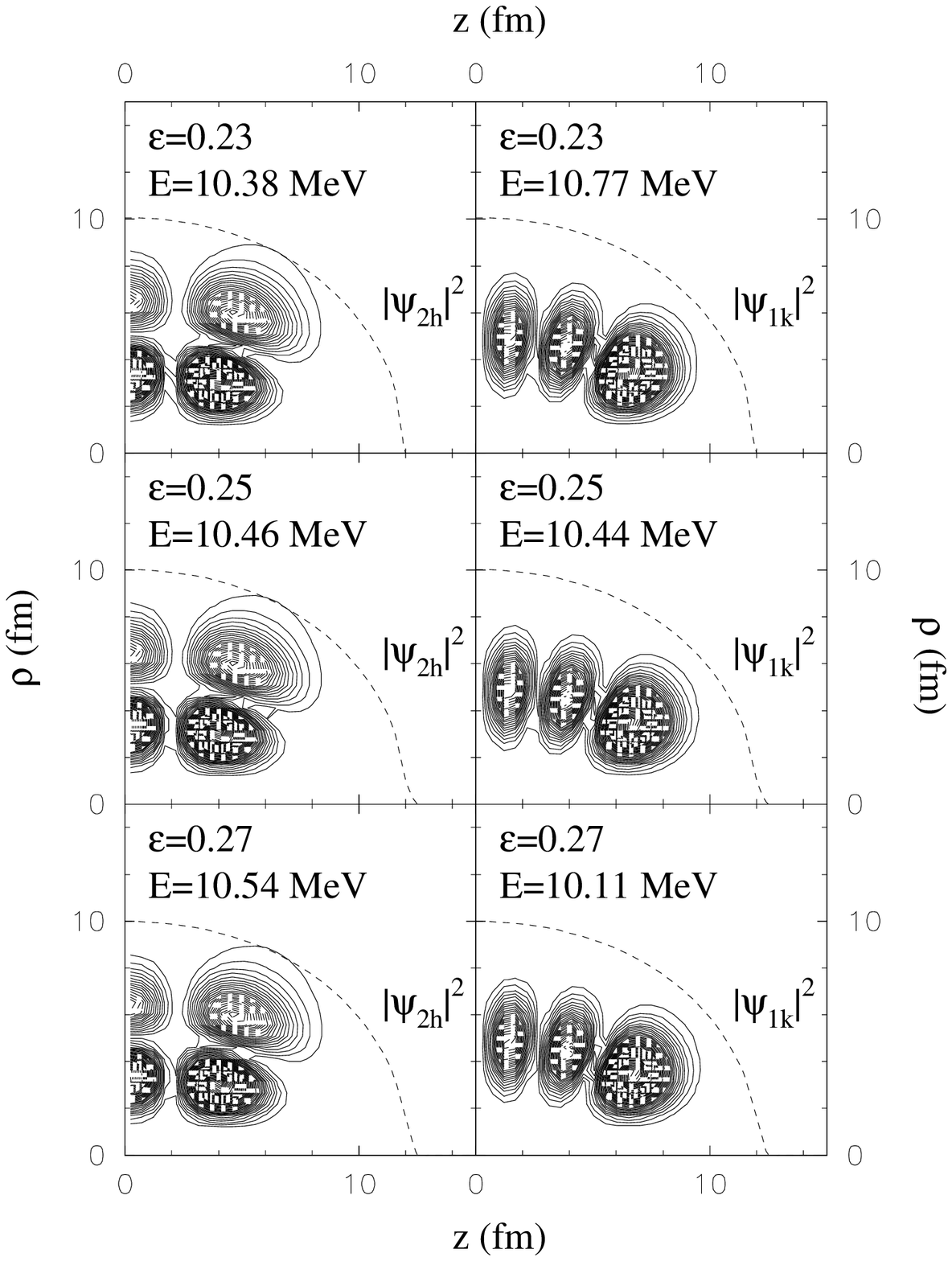,width=\columnwidth}
\caption{Square modulus of the two crossing states at and around the crossing deformation $\varepsilon_{tc}=0.25$. Dashed lines represent the deformed ridge of the potential for each deformation $\varepsilon$.}
\label{fig:states_cross}
\end{figure}

In Fig. \ref{fig:states_cross}, the square moduli of the two crossing states are plotted in the half-cylindrical plane for three different values of $\varepsilon$ at and around crossing.

\begin{figure}
\psfig{file=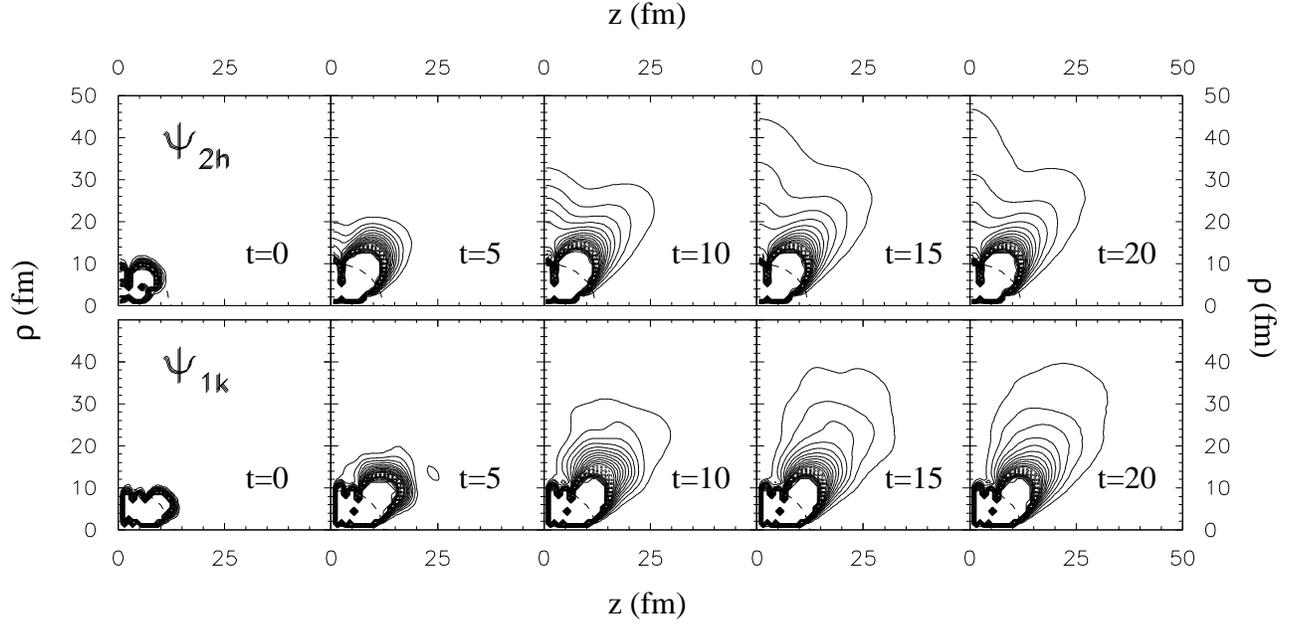,width=\columnwidth}
\caption{Time evolution of $|\psi_{2h}|^2$ and of $|\psi_{1k}|^2$ at the deformation $\varepsilon_{tc}=0.25$. The time $t$ is given in units of $10^{-22}$ sec. For clarity, contour lines with values greater than $4\times 10^{-5}$ ($3\times 10^{-6}$) in the upper (lower) row are not plotted. The potential ridge is drawn with dashed lines.}
\label{fig:stock2h}
\end{figure}

This figure shows clearly that the two wavefunctions don't mix at all during their crossing since they keep their spatial distribution constant as the potential deformation is changing. \\
Using the procedure described in Sec. \ref{sec:Formalism}, we followed the time-evolution of these two initial states for the three different deformation values $\varepsilon=0.23, 0.25, 0.27$. To illustrate the numerical results obtained, the square wavefunctions $\psi_{2h}$ and $\psi_{1k}$ (at $\varepsilon_{tc})$ have been plotted in Fig. \ref{fig:stock2h} at five different times during the calculation, showing the tunneling of these quasi-stationary states through the two-dimensional anisotropic barrier.

\begin{figure}
\psfig{file=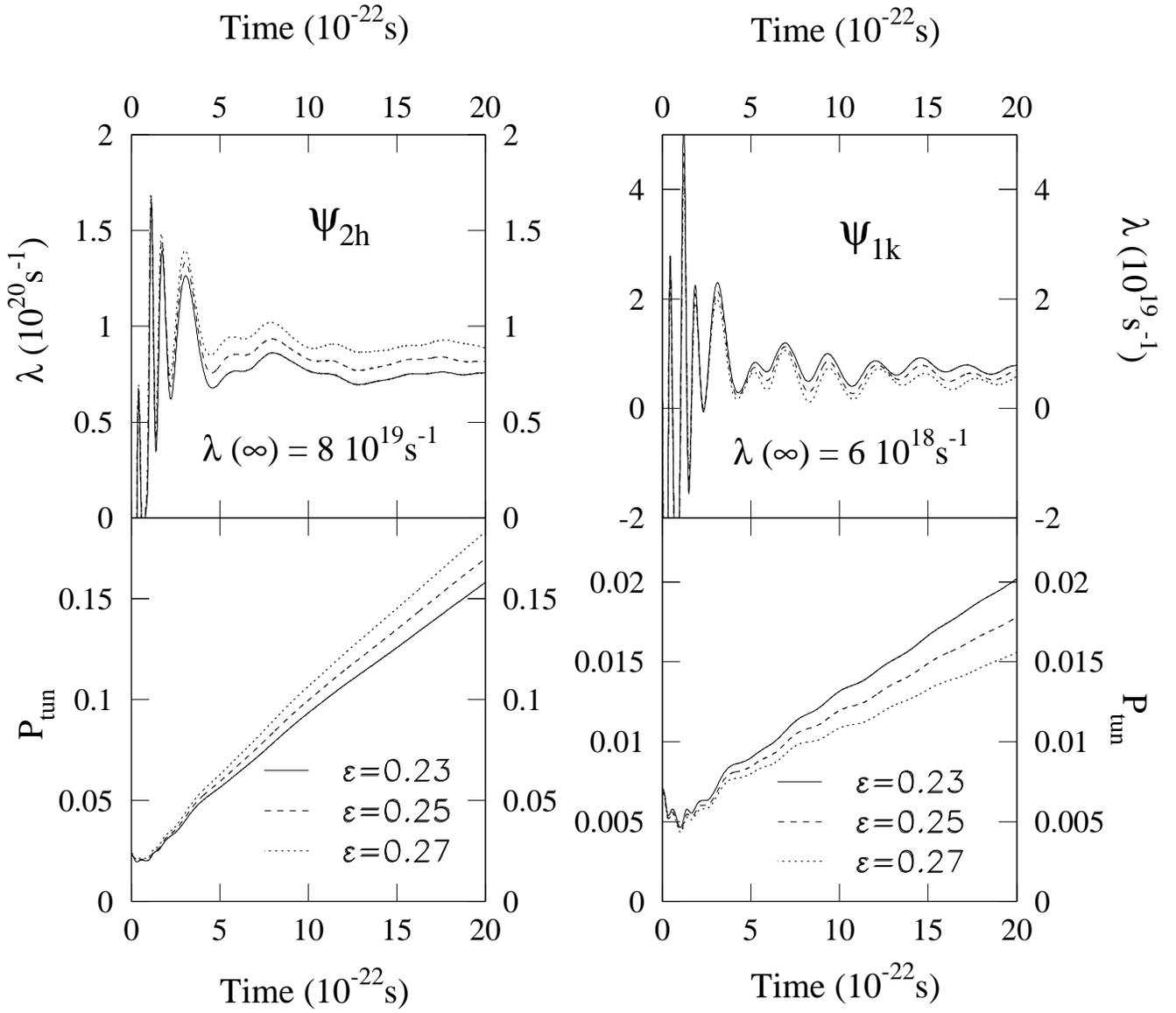,width=\columnwidth}
\caption{Time-dependent decay rates $\lambda$ and tunneling probabilities $P_{tun}$ for the two crossing states $\psi_{2h}$ and $\psi_{1k}$ obtained for three nuclear deformations. The asymptotic decay rates $\lambda (\infty)$ correspond to $\varepsilon_{tc}=0.25$.}
\label{fig:lam_cross}
\end{figure}

In Fig. \ref{fig:lam_cross} are plotted the time evolution of the decay rate as well as the tunneling probability for the two states at the three deformations considered. \\
Two different stages can be distinguished in the time evolution of $\lambda$ \cite{Ser94}. There exists an initial {\it transient time} corresponding to the ``acclimatization'' of the quasi-stationary state to its new environment and during which the decay rate $\lambda$ undergoes strong oscillations. After that, $\lambda$ is almost constant in time (in the following, this value is denoted $\lambda (\infty)$). This second stage corresponds to an exponential decay. 

From Fig. \ref{fig:lam_cross} one can extract two important results. First, for each state, the decay rate and the tunneling probability evolve very smoothly as the deformation of the potential increases. As expected, the two states don't influence each other during crossing. \\
Second, at a given deformation $\varepsilon$, the two asymptotic decay rates $\lambda (\infty)$ differ by about one order of magnitude although the two states are degenerate. In fact, this important difference can be explained by the difference in the ``residual'' angular momenta $<l>$ of the two decaying states. Although not plotted here, $<l>(t)$ is almost constant in time for both states but at two quite different values : $<l>\simeq 5.26$ for $\psi_{2h}$ and $<l>\simeq 7.56$ for $\psi_{1k}$. As shown before \cite{Tal98}, this quantity plays a major role in the determination of the decay rate.

\begin{figure}
\psfig{file=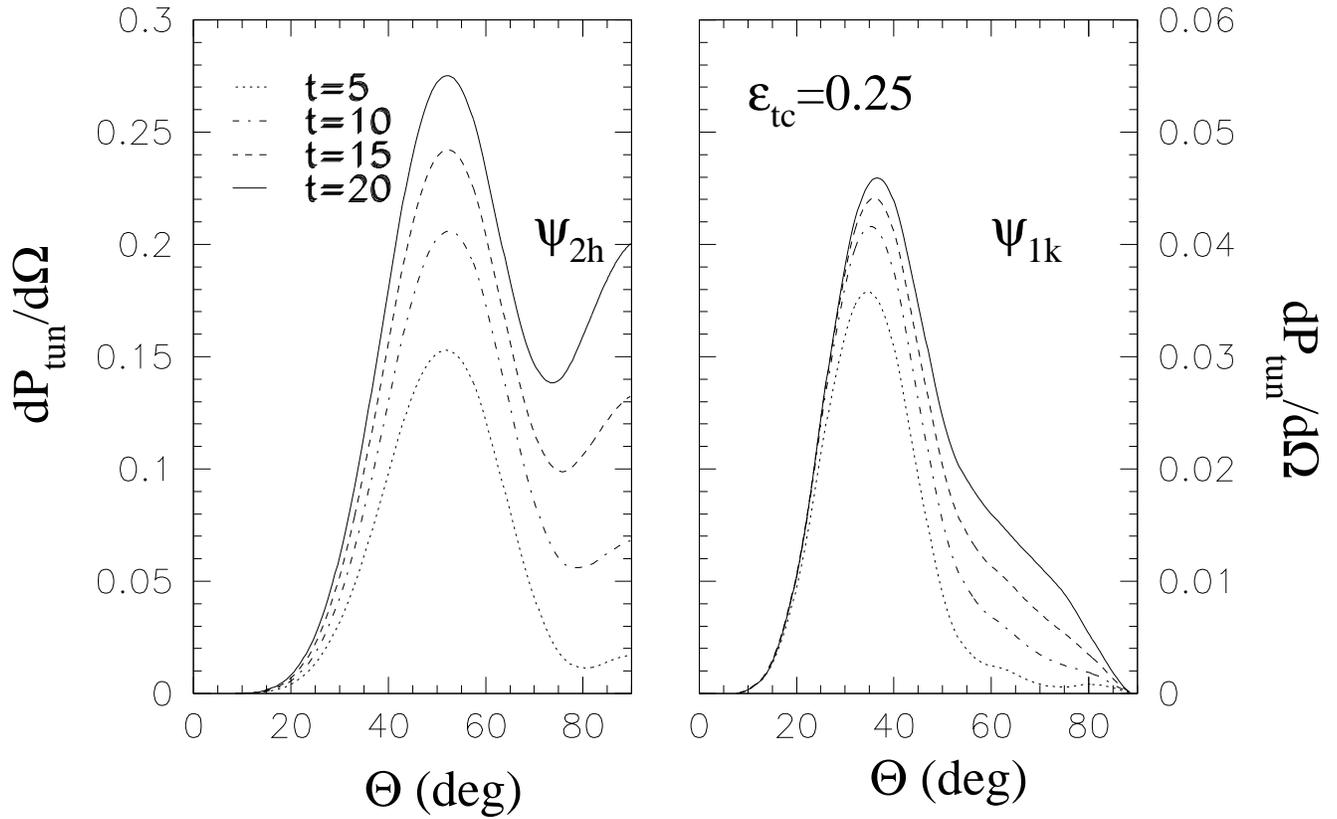,width=\columnwidth}
\caption{Time-dependent tunneling angular distributions for the two crossing states $\psi_{2h}$ and $\psi_{1k}$. The time $t$ is in units of $10^{-22}$ sec.}
\label{fig:ang_cross}
\end{figure}

From the knowledge of the wavefunctions in the cylindrical plane, we have inferred the tunneling angular distributions $P_{tun}(t,\theta)$ at different times. The results of such calculations are shown in Fig. \ref{fig:ang_cross} for the two crossing states at four different times and at the deformation $\varepsilon_{tc}=0.25$. Obviously, the two angular distributions are really different from each other and are strongly correlated to the spatial distributions of the corresponding initial quasi-stationary states. Moreover, it is worth noticing that, more generally, an emission at $\theta=90$ deg is always related to a state symmetric with respect to the $\rho$-axis (cf. $\psi_{2h}$ state). Contrarily, if the state is asymmetric (cf. $\psi_{1k}$ state), the emission is hindered at $\theta =90$ deg.

In conclusion, for a true crossing, one can determine which of the two metastable single-proton states is populated.

\subsection{Quasi-crossing states}

Now we study the influence of the mixing of two orbitals having same $\Lambda$ and $\pi$ but $\Delta N =2$ in the single-proton wave function on the tunneling behaviour.

\begin{figure}[ht]
\psfig{file=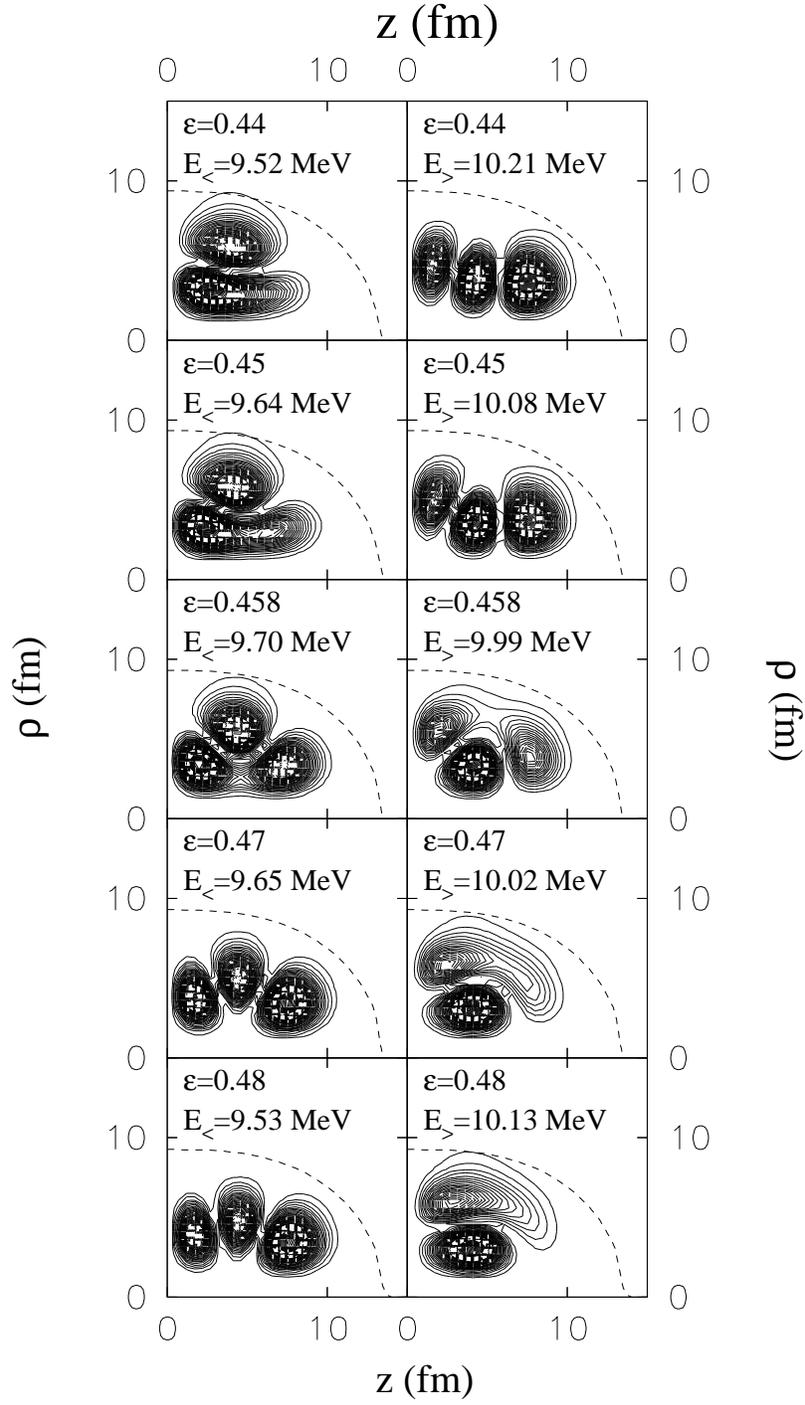,width=\columnwidth}
\caption{Contour plots of the two mixing wavefunctions at and around $\varepsilon_{qc}=0.458$. Dashed lines are the deformed ridges.}
\label{fig:states_mix}
\end{figure}

\begin{figure}[ht]
\psfig{file=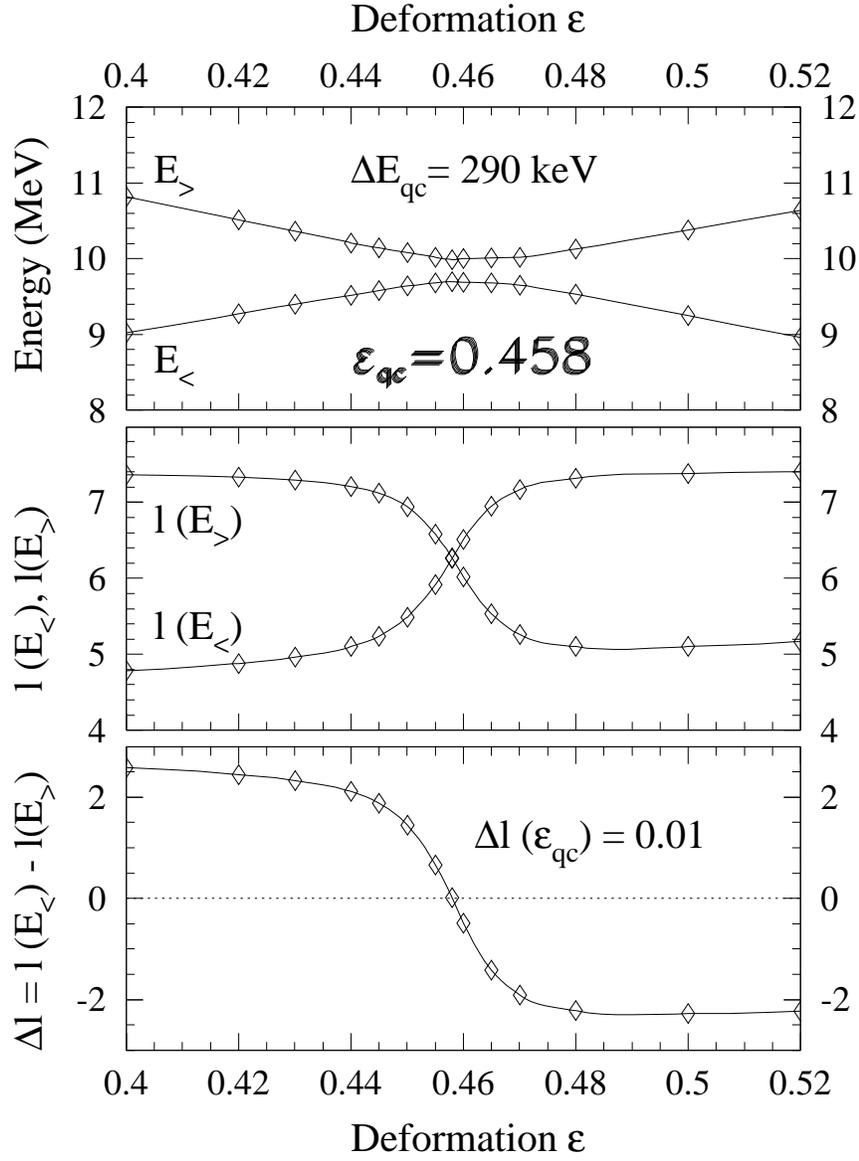,width=\columnwidth}
\caption{Average energy and ``residual'' angular momentum of the two quasi-crossing states as a function of deformation.}
\label{fig:delta_mix}
\end{figure}

Without loss of generality, we chose two states with $(\Lambda\pi)=3+$ whose levels cross at deformation $\varepsilon_{qc}\simeq 0.458$. Figure \ref{fig:states_mix} represents the square moduli of the corresponding single-particle wave functions at and around the critical value $\varepsilon_{qc}$, while the evolution of their energy and angular momentum as a function of the deformation parameter $\varepsilon$ is plotted in Fig. \ref{fig:delta_mix}. At $\varepsilon_{qc}$ the difference in energy is the smallest ($\Delta E_{qc} \simeq 290$ keV) and the two ``residual'' angular momenta are equal numerically ($\Delta <l>_{qc} = 0.01$).

Following the wave function corresponding to the lower (higher) energy $E_<$ ($E_>$) {\it vs} the deformation $\varepsilon$, one notices that its spatial distribution evolves very rapidly but continuously. Asymptotically, i.e., for $\varepsilon \gg \varepsilon_{qc}$, they have indeed interchanged their spatial distributions.

\begin{figure}
\psfig{file=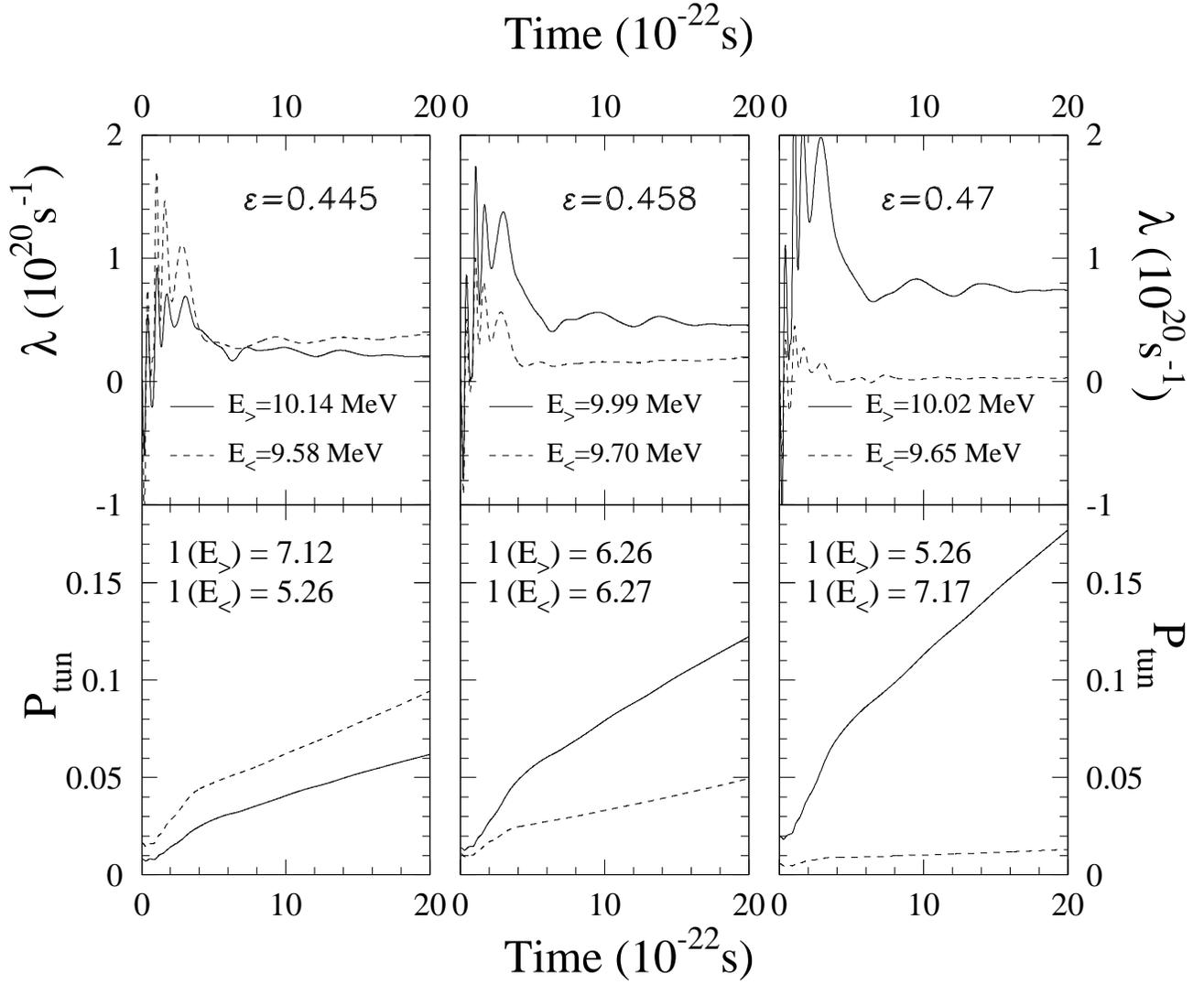,width=\columnwidth}
\caption{Time-dependent tunneling probabilities $P_{tun}$ and decay rates $\lambda$ for the two quasi-crossing states at three deformations. The values of the corresponding ``residual'' angular momenta are indicated in the lower part.}
\label{fig:lam_qcr}
\end{figure}

\begin{figure}
\psfig{file=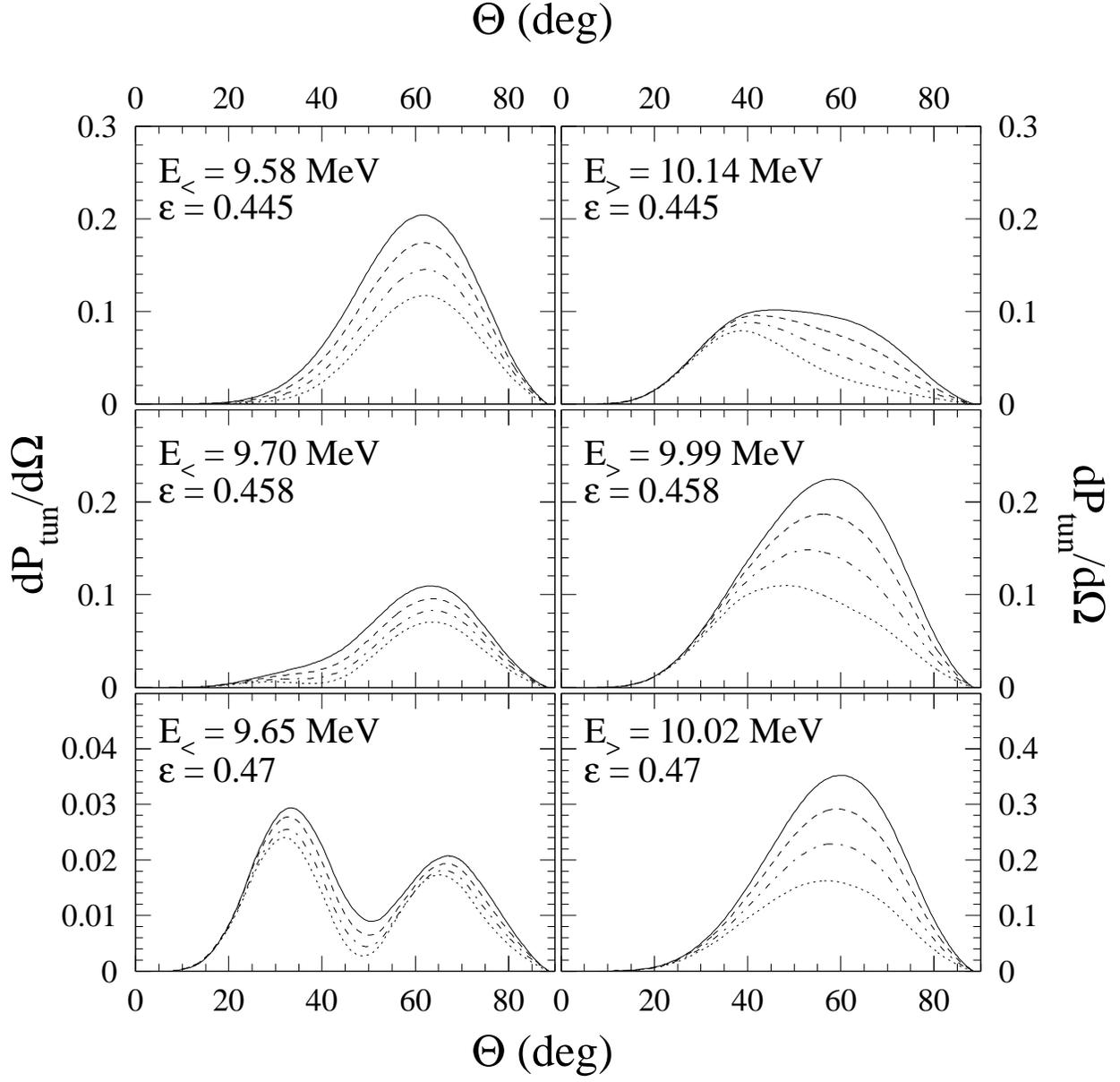,width=\columnwidth}
\caption{Time evolution of the tunneling angular distributions $dP_{tun}/d\Omega$ for the two quasi-crossing states at three deformations. The distributions have been computed at four times : $t=5, 10, 15$ and $20 \times 10^{-22}$ sec.}
\label{fig:ang_qcr}
\end{figure}

The time-dependent decay properties of these states are presented in Figs. \ref{fig:lam_qcr} and \ref{fig:ang_qcr} for three different deformations $\varepsilon$. The shapes of the angular distributions (Fig. \ref{fig:ang_qcr}) for different deformations seem to be again related to the spatial distribution of the corresponding quasi-stationary states. At the point $\varepsilon_{qc}$, the tunneling angular distributions for the states tend to coincide. \\
The variation of the decay rate from one deformation to another (Fig. \ref{fig:lam_qcr}) is clearly not correlated with the corresponding variation of the energy. However, the ratio $\lambda_{E_>}(\infty)/\lambda_{E_<}(\infty)$ (of about 3 at $\varepsilon_{qc}$) can be explained by the difference in energy ($E_>-E_<=290$ keV) using stationary one-dimensional WKB estimates.

\begin{figure}[htbp]
\psfig{file=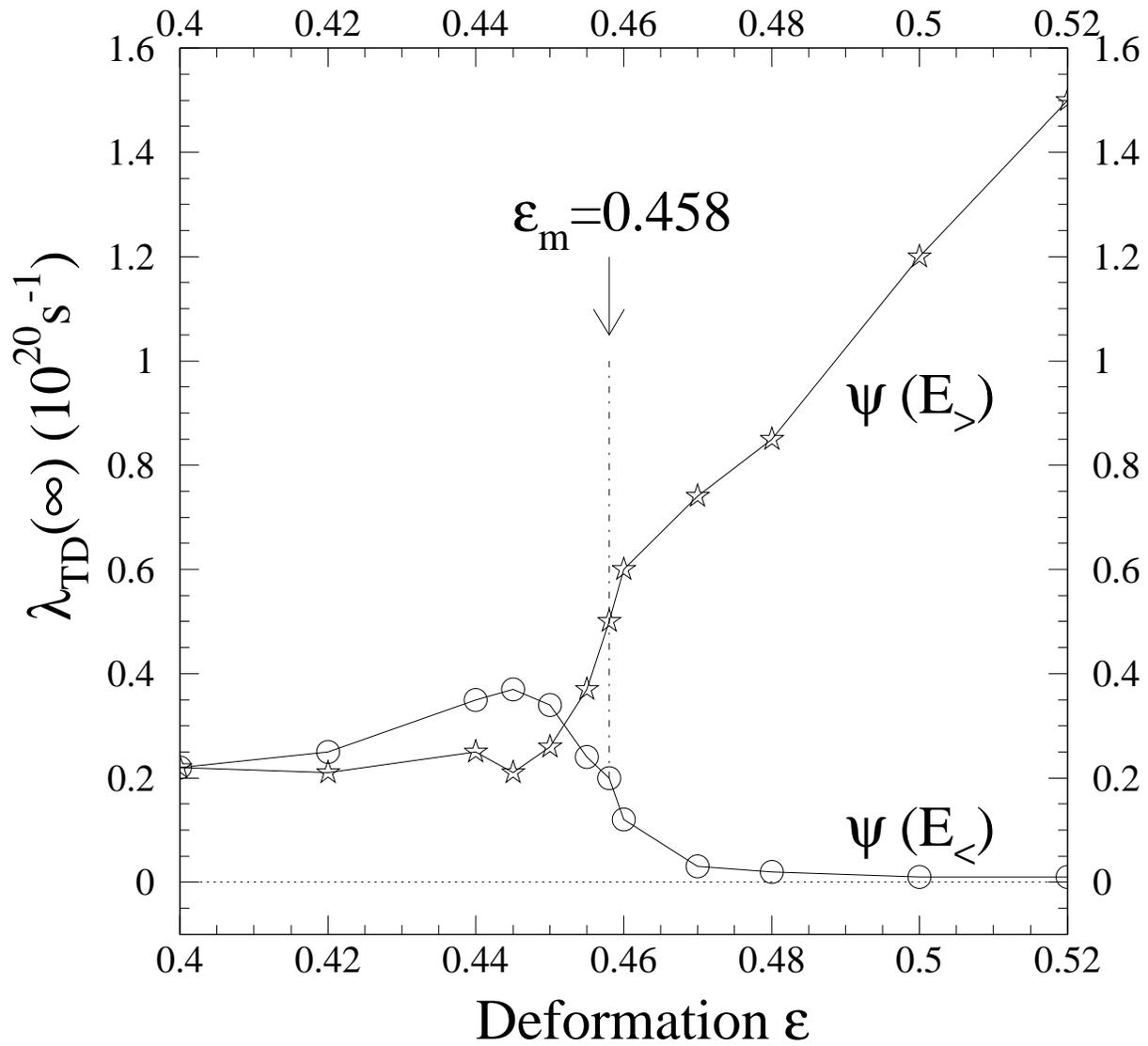,width=\columnwidth}
\caption{Asymptotic decay rates $\lambda (\infty)$ for the two initial $(3+)$ states {\it vs} the deformation $\varepsilon$.}
\label{fig:laminfty_mix}
\end{figure}

To study in more detail the dependence of $\lambda$ on nuclear deformation, time-dependent calculations have been performed for the same quasi-stationary states at several deformations on both sides of $\varepsilon_{qc}$. The asymptotic values of the decay rates $\lambda (\infty)$ have been plotted in Fig. \ref{fig:laminfty_mix} as a function of the deformation $\varepsilon$. It is clearly shown that at the quasi-crossing point ($\varepsilon=\varepsilon_{qc}$), the two asymptotic decay rates are well separated. Contrarily, at $\varepsilon =0.4$, the two $\lambda (\infty)$ are almost equal. This effect can be understood in terms of the competition between the energy of the decaying state and its ``residual'' angular momentum value $<l>$. For $\varepsilon < \varepsilon_{qc}$, the higher energy state has the higher $<l>$ value. On the other hand, for $\varepsilon > \varepsilon_{qc}$, the higher energy state corresponds to the smaller $<l>$ value. Both effects act in the same direction, strongly enhancing the decay rate.

In conclusion, for a quasi-crossing, one can determine the energy splitting and whether the proton has jumped into the upper level or not (Landau-Zener effect).

\section{Summary}

The two-dimensional Time-Dependent Schr\"odinger Equation has been solved numerically for prepared initial single-particle quasi-stationary states decaying through an anisotropic, non separable, potential barrier. The special cases of true crossing and quasi-crossing have been studied within this approach. Time-dependence of quantities like tunneling probability, decay rate and tunneling angular distributions have been calculated. \\
It has been shown that two (accidentally) degenerate proton states decay very differently due to their different spatial distribution and ``residual'' angular momentum. Hence, one-dimensional semi-classical approximations of their decay rate are not expected to be reliable. \\
The mixing of two orbitals having the same main quantum numbers in the single-proton wave function strongly influences the decay behaviour. At the quasi-crossing deformation, the difference between the proton decay rates from the two states involved can be entirely explained by the difference in their energy. A nearby deformation, where these two rates are the same, can always be found due to the combined effects of the residual angular momentum and of the energy of the decaying states.

Proton decay in strongly deformed nuclei has been recently observed \cite{Dav98}. Extensions of this type of work would allow a detailed study ot the level-crossing phenomenon, including the structure of the nuclear states involved.

Further developments of the present work should include generalization to proton decay through a time-dependent potential barrier since only a dynamical approach could tackle such a problem. There are indeed cases in which the parent and daughter nuclei have very different deformations, e.g., prolate-oblate transitions or emission during nuclear fission.


\bibliographystyle{plain}

\newpage
\onecolumn

\end{document}